\documentclass[10pt,letterpaper]{article}

\usepackage[top=0.85in,left=2.75in,footskip=0.75in]{geometry}

\usepackage{amsmath,amssymb}

\usepackage{listings}             

\usepackage{siunitx}

\usepackage{changepage}

\usepackage[utf8]{inputenc}
\usepackage{soul}

\usepackage{textcomp,marvosym}

\usepackage[utf8]{inputenc}
\usepackage[english]{babel}
 
\usepackage[
backend=biber,
style=alphabetic,
citestyle=authoryear
]{biblatex}
\usepackage{nameref,hyperref}

\usepackage{microtype}
\DisableLigatures[f]{encoding = *, family = * }

\usepackage[table]{xcolor}

\usepackage{array}

\newcolumntype{+}{!{\vrule width 2pt}}

\newlength\savedwidth

\newcommand\thickhline{\noalign{\global\savedwidth\arrayrulewidth\global\arrayrulewidth 2pt}%
\hline
\noalign{\global\arrayrulewidth\savedwidth}}

\raggedright
\usepackage[aboveskip=1pt,labelfont=bf,labelsep=period,justification=raggedright,singlelinecheck=off]{caption}

\usepackage{lastpage,fancyhdr,graphicx}
\usepackage{epstopdf}
\fancyheadoffset[L]{2.25in}
\fancyfootoffset[L]{2.25in}
\begin{document}
\vspace*{0.2in}

\begin{flushleft}
{\Large
\textbf\newline{Effects of publication bias on conservation planning} 
}
\newline
\\
\hypersetup{
pdfauthor={Raffael Hickisch, Timothy Hodgetts, Paul J. Johnson, Claudio Sillero, Klement Tockner, David W. Macdonald},
pdftitle={Sample Bias Compromises Conservation Priorities},
pdfkeywords={accessibility,geographic bias, maps, publication density, sampling effort}
}

Raffael Hickisch\textsuperscript{1*},
Timothy Hodgetts\textsuperscript{1},
Paul J. Johnson\textsuperscript{1},
Claudio Sillero\textsuperscript{1,2},
Klement Tockner\textsuperscript{3,4,5},
David W. Macdonald\textsuperscript{1}
\\
\bigskip
\textbf{1} Wildlife Conservation Research Unit (WildCRU), Zoology, University of Oxford. The Recanati-Kaplan Centre, Tubney House, Tubney OX13 5QL, United Kingdom
\\

\textbf{2} Born Free Foundation, Horsham  RH12 4QP, United Kingdom
\\
\textbf{3} Leibniz-Institute of Freshwater Ecology and Inland Fisheries (IGB), 12587 Berlin, Germany
\\
\textbf{4} Austrian Science Fund (FWF), 1090 Vienna, Austria
\\
\textbf{5} Institute of Biology, Freie Universität Berlin, 14195 Berlin, Germany
\\
\bigskip

* raffaelhickisch@gmail.com

\bigskip

\hl{\textbf{Note:} This is a pre-peer reviewed version as per Jan-2018; it has been accepted in Conservation Biology on 8 April, 2019 with the DOI:10.1111/cobi.13326. This article may be used for non-commercial purposes in accordance with Wiley Terms and Conditions for Use of Self-Archived Versions.}

\end{flushleft}
\section*{Abstract}
Conservation planning needs reliable information on spatial patterns of
biodiversity. However, existing data sets are skewed: some habitats,
taxa, and locations are under-represented. Here, we map geographic
publication density at the sub-national scale of individual 'provinces'.
We query the Web of Science catalogues SCI and SSCI for
biodiversity-related publications including country and province names
(for the period 1993-2016). We combine these data with other
provincial-scale factors hypothesised to affect research (i.e. economic
development, human presence, infrastructure and remoteness). We show
that sites that appear to be understudied, compared with the
biodiversity expected from their bioclimatic conditions, are likely to
have been inaccessible to researchers for a diversity of reasons amongst
which current or recent armed conflicts are notable. Finally, we create
a priority list of provinces where geographic publication bias is of
most concern, and discuss how our provincial-scale model can assist in
adjusting for publication biases in conservation planning.



\section*{Introduction}
National and international conservation policies rely on global
biodiversity data sets. Repositories such as the IUCN Red List and the
Global Biodiversity Information Facility (GBIF) synthesise existing data
sets to support conservation planning. However, knowledge of the
abundance and distribution of biodiversity remains patchy. Even the
global number of species is unknown, and the spatial distribution of
those species that have been described is often uncertain -- knowledge
deficits termed the ``Linnaean'' and ``Wallacean'' shortfalls,
respectively (\cite{brito_overcoming_2010}). Hence, there are continuing calls to improve
biodiversity information (\cite{jetz_integrating_2012}). Indeed, it would be
informative to have `maps of ignorance' that map and quantify present
knowledge deficits (\cite{rocchini_accounting_2011}). Concurrently, open
access publishing policies and science communication strategies to
redress the underrepresentation of less affluent regions need reform
(\cite{wilson_conservation_2016}), such as better targeted research
priorities.

The spatial patchiness of existing biodiversity data sets reflects
historic and contemporary research efforts. Hence, these data sets
incorporate biases, emphases and omissions (e.g. \cite{boakes_distorted_2010}). Research efforts have tended to focus on favoured taxa and
habitats, such as vertebrates and forests (\cite{fazey_what_2005}).
Within taxonomic groups further biases exist. For example, research on
felids is biased towards those with a large body size (\cite{brodie_is_2009})
and \cite{brooke_correlates_2014} found that body size, range size and diet
explain uneven patterns of research efforts across carnivores.

Uneven spatial patterns occur for a variety of reasons. Accessibility of
is a strong influence on the likelihood of a location being researched.
\cite{reddy_geographical_2003}, for example, demonstrated that sampling
locations of passerine birds in Sub Saharan Africa were clustered near
roads, along rivers, and around urban areas. The practicalities faced by
researchers striving to adapt to living and working conditions are
amongst the components of accessibility. Socio-economic conditions (e.g.
levels of affluence, infrastructure, language, security) privilege
specific areas for research, and thus influence spatial sampling
patterns (\cite{martin_mapping_2012,amano_four_2013,meyer_range_2016}). 

For example, \cite{fisher_global_2011} showed that
research efforts on coral reefs were positively correlated with an
area's per capita gross domestic product. Research effort is also known
to cluster around areas that have previously been deemed of interest.
This might be considered a consequence of their already documented
biodiversity and consequently known threats to the biodiversity of these
areas, but may also be due to the practical and infrastructural benefits
provided by biological field stations and established monitoring sites
(see \cite{tydecks_biological_2016}). Thus, \cite{martin_mapping_2012}
demonstrated that protected areas are over-represented, as are sites in
temperate deciduous woodlands and wealthy countries. Whatever the
causes, these biases can lead to a self-reinforcing cycle whereby
`funding begets biodiversity' (\cite{ahrends_funding_2011}).

The geographic sampling biases in biodiversity research have
implications for macro-ecological modelling (\cite{yang_geographical_2013}),
conservation prioritisation \cite{de_ornellas_impact_2011}, and
governance audits like those of the Intergovernmental Science-Policy
Platform on Biodiversity and Ecosystem Services (IPBES). While modellers
have developed methods to correct for sampling biases in species range
modelling (\cite{kramer-schadt_importance_2013,engemann_limited_2015})
and in applying systematic conservation planning tools (\cite{rondinini_tradeoffs_2006}), such remedies cannot always address the persistent
undersampling of biodiversity (\cite{wilson_conservation_2016}). For example,
(\cite{grenyer_global_2006}) demonstrated little spatial congruence
between the occurrence of threatened bird and mammal species (or across
these taxa) at the spatial scales relevant for conservation planning. In
addition, re-directing effort from well-sampled towards under-sampled
locations may improve the efficiency of conservation actions (\cite{aranda_designing_2011}). 

\cite{sastre_taxonomist_2009} demonstrated that, under
some plausible assumptions, if the aim of a study is to describe the
spatial distribution of species richness in a given area, then surveying
under-sampled locations is more effective than surveying areas known to
be of high species richness. \cite{de_ornellas_impact_2011} used a
simulation study to find out that focussing on poorly surveyed areas
maximises the benefits of surveys for conservation planning.

We developed a global, provincial-scale map of biodiversity publication
density to identify areas where low accessibility is associated with
research deficits. We define `provinces' as sub-national administrative
units, as recognised in the Global Administrative Areas database
(\cite{gadm_gadm_nodate}), covering differently named areas (e.g.,
provinces, departments, counties) of various size. For each province, we
mapped the number of biodiversity research publications by area using
the Web of Science catalogues SCI and SSCI. This map can be considered a
surrogate for sampling effort, as demonstrated by \cite{lobo_database_2008}. We
modelled the associations between factors affecting accessibility (i.e.,
remoteness, infrastructure, and development) and publication density,
identifying provinces where the shortfall in publications is linked to
accessibility, but where human impact on biodiversity is also very
likely. We suggest that such a `map of ignorance' can facilitate a
prioritisation of future biodiversity research that adjusts for current
bias.

\section*{Methods}
For each province we queried the Web of Science (WoS) catalogues SCI and
SSCI for publications labelled in the \emph{``Biodiversity \&
Conservation''} category (covering publications appearing in any journal
listed under that category; not to be confused with the journal of the
same name) that included the country name (in multiple languages and
writing styles) as well as the province name. We limited our search to
the period 1993 to 2016 -- starting with the United Nations Earth Summit
conference in Rio de Janeiro, to capture a period of sustained global
research in biodiversity. Many more biodiversity publications were
identifiable to country level (n= 1,229,636) than to province level (n =
54,144). As some provinces had no results for the biodiversity and
conservation WoS category, we inferred values for the number of
publications based on the ratio of biodiversity research to overall
research within a country. Assuming that this proportion was constant
for a country we multiplied the province's overall total number of
publications by this proportion and adjusted the result by province area
(n= 44,858 publications on province level, see Supplementary Information
(SI) for details). We refer to this hereafter as `inferred publication
density', with publications per km\textsuperscript{2} as the unit.

The queries were performed using the R-package \emph{rwos} (\cite{barnier_r_2019}) and the Web of Science Lite API~(\cite{reuters_web_2012});~\emph{country}
and~\emph{province} name were taken from the~GADM data set~(\cite{gadm_gadm_nodate}) and the R-package~\emph{countrycode}~(\cite{arel-bundock_r_2019}) and
results were filtered in the~different Web of Science domains
(\emph{all} = no constraint,~\emph{cons} = biodiversity and conservation
as single category). The map of inferred biodiversity publication
density was produced in R-language 3.1 with the script
in~\cite{raffael_r_2017_rwosconsindex}.

We hypothesised that the publication density in a province would be
affected by the degree of accessibility of that province for
researchers. Given that there were numerous potential drivers affecting
both accessibility and biodiversity, we used an ordination technique to
reduce the data to a small set of orthogonal axes summarising the input
variables. We conducted a Principal Components Analysis (PCA) using the
\emph{prcomp} function in R. We used two proxies of \emph{ecological
conditions}: mean precipitation and temperature data derived from
Worldclim (\cite{fick_worldclim_2017}); two measures of \emph{human presence}:
one based on the NASA night-time light data set corrected for gas flares
(\cite{ngdc_version_2013}) and the other derived from weekly fire clusters of the
NASA VIIRs data set~(\cite{nasa_viirs_nodate}). Within the data on fire clusters we
corrected for large bushfires by spatially clustering reported fire
points by week (see SI; weekly fire clusters may be a better indicator
than simple fire point counts, as these remove the effects of large bush
fires that can ignite naturally or result from a one-time human action).
These fire indices complement those quantifying human footprint insofar
as the latter fail to capture the impact of nomadic agro-pastoral
communities. We also developed a \emph{remoteness} \emph{index} based on
mean travel time to major cities, capitals, airports and seaports
(\cite{nelson_travel_2008,giraud_shortest_2019}) and a \emph{world road area} index (derived
from~\cite{ibisch_global_2016}). We used mean life expectancy at birth
(from 1966 to 2016) as a measure of \emph{human development} as well as
an index of the relative living conditions in a province (\cite{world_bank_group_life_nodate}). We cut all geo-spatial data to the extent of the
province shape, and calculated~mean values (for raster data), as well as
the size (in m\textsuperscript{2}) and counts for all vector data sets.
These data are stored on file per GADM province record (see SI \ref{table6} for
further details). We had to remove 53 of 204 countries, and 184 of 3,417
provinces prior to the PCA, due to missing data. In most cases, the
missing data related to life expectancy and applied to small island
groups. These are often not listed separately in the relevant World Bank
data sets. We log-transformed the variables with large variance and high
skew (i.e., publication density, road area, fire clusters, and
remoteness), and corrected for province area. We plotted location of
provinces as a function of PCA 1 and PCA 2 scores using the
\emph{ggfortify} R-package (\cite{tang_ggfortify:_2016}).

We fitted a linear model for the subset of provinces where the number of
publications was known, and used inferred publication density as the
response and actual conservation publications as predictor. We also
fitted a model for the same subset with the remoteness index as
response, and night light intensity, road and province area measures as
predictor. A third model used the fire clusters as response and the
night lights and WDPA Category I and II protected areas as predictors.
Finally, we ran a linear mixed effects model (LMER) with inferred
publication density as the response, and the above variables as
predictors, correcting for country as a random effect, and also
incorporating protected areas and IUCN terrestrial mammal ranges (\cite{iucn_iucn_nodate}) to assess which predictors best explain observed patterns in
biodiversity publication density (taking into account any pattern of
collinearity among them). Full details of all provinces are available
online and in the Appendix (see Table  \ref{table5}) and online (\cite{hickisch_data_nodate}).

We identified priority provinces, where remoteness was most strongly
linked to publication density and where climate (and latitude) indicate
that biodiversity is likely to be high. We did so by summing the
absolute axis scores of the synthetic component PC1, and the synthetic
component PC2 for those provinces in the quadrant most characterised by
remoteness and high number of fire clusters.

\begin{figure}[!h]
\begin{adjustwidth}{-2.25in}{0in} 
\centering
\includegraphics[width=\linewidth]{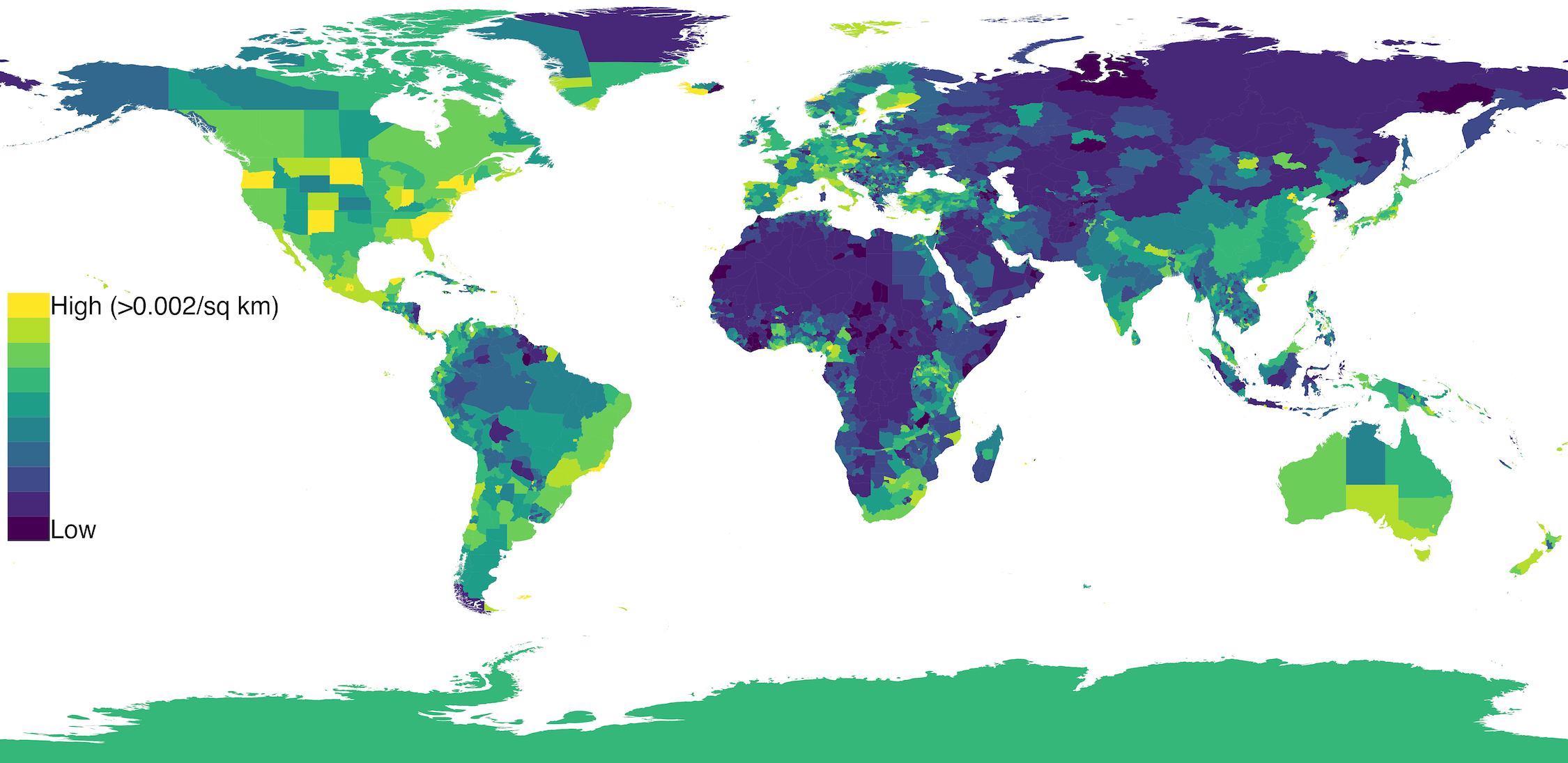}
\caption{{\bf Map of publication density.}
Map of density of publications of the biodiversity and
conservation category in the Web of Science SCI and SSCI catalogues that
mention country and province name (recording period: 1993 to 2016),
adjusted for the geographic size of the provinces (km\textsuperscript{2}). The data for this map is available at \cite{hickisch_aggregated_2017}; An interactive
version of this map is available at \href{https://bit.ly/publication_density_map}{bit.ly/publication\_density\_map}}
\label{fig1}
\end{adjustwidth}
\end{figure}

\begin{figure}[!h]
\includegraphics[width=\linewidth]{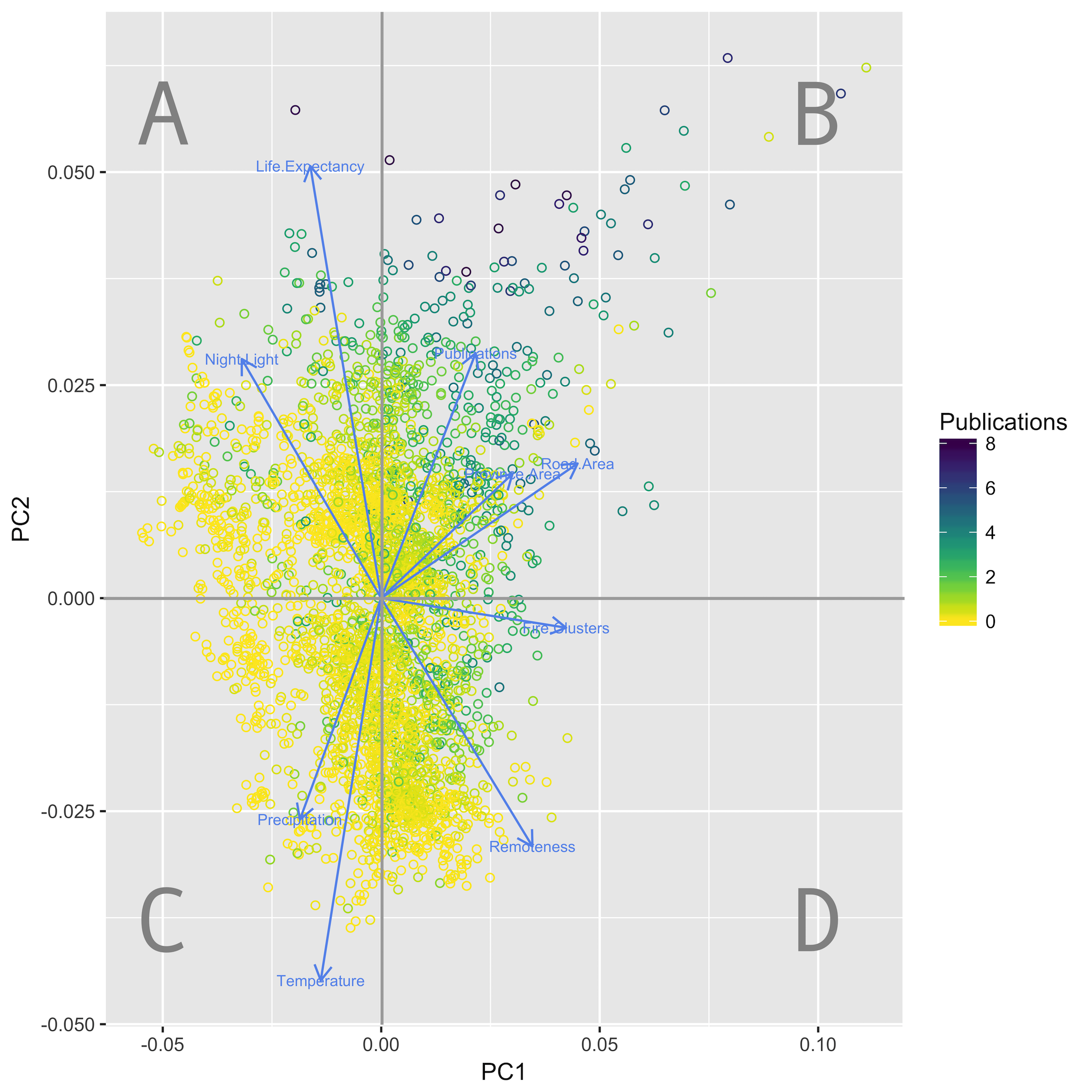}
\caption{{\bf Principal Component Analysis plot PC1 PC2.}
Dots represent provinces, and color indicates log transformed biodiversity and conservation related publication density (data available at \cite{hickisch_supplementary_2017}).
The tropical provinces (high temperature and precipitation) that are
less affluent (life expectancy) show low economic activity (nightlight),
are remote (remoteness), and exhibit high human presence (fire
clusters). In general, these provinces appear less studied -- even
though they are often located adjacent to known biodiversity hotspots.}
\label{fig2}
\end{figure}

\begin{figure}[!h]
\begin{adjustwidth}{-2.25in}{0in} 
\centering
\includegraphics[width=\linewidth]{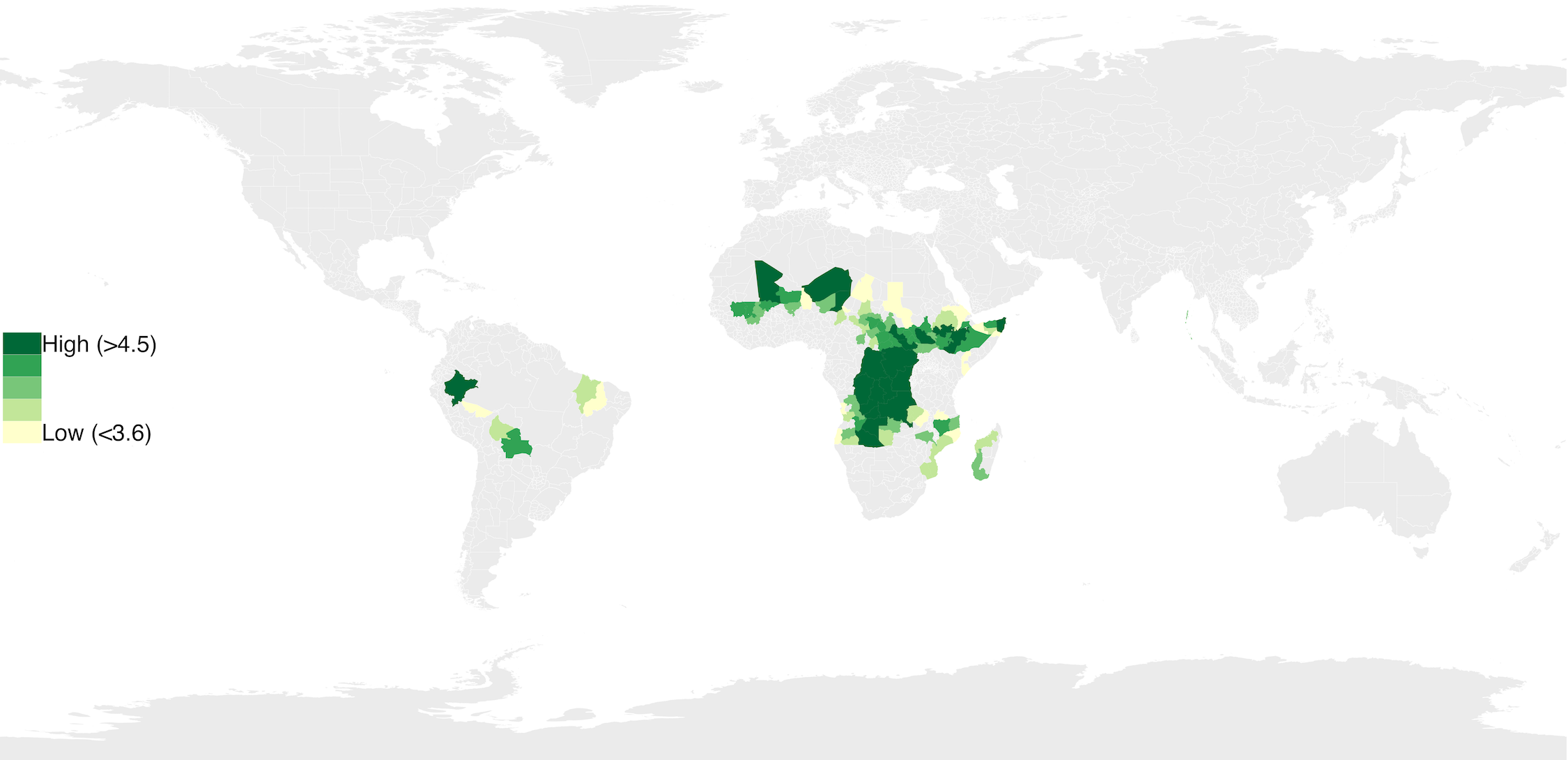}
\caption{{\bf 100 Priority Provinces.}
The location of 100 priority provinces where the accessibility
bias in geographic reporting bias is most apparent. Provinces of
Quadrant C in the PCA (see Fig. 2) are ranked by their combined
coordinate score (PC1 positively and PC2 negatively) and indicated by
color (data at \cite{hickisch_supplementary_2017}}
\label{fig3}
\end{adjustwidth}
\end{figure}

\begin{table}[!ht]
\begin{adjustwidth}{-2.25in}{0in} 
\centering
\caption{
{\bf Results of linear mixed model with the number of
publications in the Web of Science biodiversity and conservation
category as response, and predictors used in glmm model.}}

\begin{tabular}{@{}llllll@{}}

\textbf{Response} & \textbf{Predictor} & \textbf{Level} & \textbf{Estimate} & \textbf{SE} & \textbf{p-value} \\ \thickhline
Publications & Number.ProtectedAreas & Province & 7.53E-02 & 1.31E-02 &
8.65E-09\\
& Number.TerrestrialMammals & Province & 8.14E-02 & 2.21E-02 &
2.37E-04\\
& ProtectedArea.Area.Cat.IabII & Province & 7.42E-02 & 1.16E-02 &
1.91E-10\\
& Night.Light & Province & 2.10E-01 & 1.48E-02 & 2.00E-16\\
& Life.Expectancy & Country & 1.21E-01 & 4.94E-02 &
1.53E-02\\
& log(Road.Area) & Province & 2.79E-01 & 2.34E-02 &
2.00E-16\\
\end{tabular}

\begin{flushleft} Results of linear mixed model with the number of
publications in the Web of Science biodiversity and conservation
category as response, and predictors used in glmm model, mainly derived
from spatial data sets (3428 observations). Life Expectancy is available
only at the country level (n=204), hence country was used as grouping
variable.
\end{flushleft}
\label{table1}
\end{adjustwidth}
\end{table}

\section*{Results}

The highest density of publications in the biodiversity and conservation
WoS category were observed in more affluent countries and those with
known biodiversity hotspots (Fig.~\ref{fig1}). As expected from the latitudinal
biodiversity gradient theory, high latitude areas were associated with
lower publication density. At the same time, publication density was low
for some areas close to the equator, such as Central Africa, too.

There was a positive and significant correlation between the number of
biodiversity and conservation publications at province level and the
inferred number of publications (R\textsuperscript{2} = 98.4\%), based
on the assumption that the number is proportional to the size of the
province; the inferred number predicted almost all the variation in the
real number.

The remoteness index increased with the area of a province, and
decreased with night light level and road density (total
R\textsuperscript{2} = 45.0\%, model details in SI). The fire cluster
index decreased with increasing electrified areas (night light
intensity) and with increasing protected areas (total
R\textsuperscript{2} = 61.0\%).

The number of inferred publications of a province correlated most
strongly with the country's inhabitants' life expectancy at birth, night
light intensity, and log-transformed road area (see Table~\ref{table1}). The number
of biodiversity and conservation publications associated with a province
increased with the indices of human development scores and road area
(see Table 1). There was also a positive correlation between publication
density and the number of protected areas in the respective province
(see Table 1), as well as with the number of terrestrial mammal species
present in a province according to IUCN distribution range maps.

The PCA (Fig.~\ref{fig2}) demonstrates correlations between research deficits
(low numbers of publications) and accessibility (i.e. lower than mean
values of life expectancy, night light, road area, and remoteness). The
first two components together explained 54.1\% of the variance in the
input variables (see SI). Visually, four distinct groups of provinces
could be defined, corresponding to the quadrants of the PCA (A -- D; see
Fig.~\ref{fig2}). Quadrant A encompassed provinces characterised by high levels
of human development (i.e. high life expectancy) and high economic
activity (in terms of night light). In addition, these provinces
exhibited a high number of biodiversity publications as well as low
temperature and low rainfall, compared to provinces in Quadrants C and
D. Quadrant B included provinces characterised by high number of
biodiversity publications, high density of roads and large area; again
with relatively low temperature compared to provinces Quadrants C and D.
Quadrant C included primarily tropical provinces, characterised by high
temperature and precipitation, with low number of biodiversity
publications (relative to Quadrants A and B), and lower degree of
remoteness than those on Quadrants B and D. Accessibility issues may be
of less relative importance in provinces located in Quadrant C. Finally,
Quadrant D contains provinces that exhibit a low number of biodiversity
publications, but high levels of remoteness (long travel times to major
cities, capitals, seaports, and airports) and distinct human presence.
At the same time, these provinces are located in less affluent regions,
characterised more by fire clusters rather than by road area and night
light.

Fig. ~\ref{fig3} shows the scores of the `top 100' priority provinces (largely
within 15 degrees latitude to the equator) for which the factors in
Quadrant D (Fig.~\ref{fig2}) where the accessibility bias in geographic
biodiversity reporting is likely to be of most influential (full list of
``priority provinces'': \ref{table9}).

\section*{Discussion}
In this paper, we identified provinces with marked differences between
the actual density of publications concerning biodiversity and
conservation and the expected density of publications based on
bioclimatic conditions and the latitudinal diversity gradient ( \cite{gaston_global_2000,hillebrand_generality_2004,mannion_latitudinal_2014}). Differences between
actual and expected publication density are most likely to be related to
the degree of accessibility of provinces. By integrating a large array
of available data sets, the methodology presented here may provide a
useful tool in optimizing future biodiversity monitoring and research
efforts, regionally and globally.

Here we discuss the choice of complementary geospatial indicators. Based
on recent case studies from Central Africa we moreover exemplify what a
blurred view on the state of biodiversity in some provinces can imply.
In addition we illustrate how an improved understanding of situation in
100 priority regions can make the management of well-known biodiversity
hotspots more resilient.

Conventional indicators of the human footprint, such as the occurrence
of roads, villages, and an index of night light, may underestimate the
actual human impact in some provinces. For example, using fire clusters
as a complementary indicator will allow assessing the effects of
pastoralism and land clearance for farming, which have remained
invisible for long periods. This indicator reveals an additional
category of under-studied provinces located in Quadrant D (Fig. 2),
which are of particular concern for current and future conservation
efforts. These provinces are expected to exhibit high biodiversity.
However, road infrastructure is poor and security may be limited,
therefore knowledge about actual biodiversity status may be limited too
(see \cite{reddy_geographical_2003}).

Most biological field stations are located in or close to protected
areas. Concurrently, many of these areas are protected due to their high
biodiversity, which again creates a positive accessibility bias towards
these areas (\cite{tydecks_biological_2016}). In addition, funding
opportunities are most likely higher for these areas, compared to
unprotected 'biodiversity hotspots' (\cite{ahrends_funding_2011}).
Similarly, the positive correlation between publication density and
terrestrial mammals' distribution ranges can be interpreted as
reflecting appropriate targeting of the areas of most need. Furthermore,
species distribution range maps are frequently the outcome of biased
reporting efforts. For example, range contraction estimates of the
African lion in Central Africa are based on only very recent field
observations (H. Bauer, personal communication, August 30, 2017;
\cite{brugiere_large-scale_2015}).

The 100 priority provinces exhibiting distinct gaps in
biodiversity-related research, primarily due to low accessibility, share
common attributes (see Fig. 3). These provinces are not necessarily very
rich in biodiversity; however, they are located in remote areas, exhibit
low human development index scores, and contain poor infrastructure.
Most priority provinces are located in Central Africa, central Western
Africa, Bolivia and the Amazonian regions (Peru and lowland areas of NE
Brazil), as well as in the Andaman and Nicobar islands (India; see Fig. ~\ref{fig3}).

The inclusion of fire clusters in the model reflects the intensity of
human impact in those provinces where standard indicators (i.e., night
light, road area) would not detect human impacts. Therefore, our list of
priority provinces considers `potential threat'; in addition to actual
research deficits. Indeed, we did not highlight provinces that have low
publication density \emph{and} low human presence (as measured by fire
clusters) as biodiversity in these areas is presumably under a lower
level of threat - although these provinces may exhibit important
knowledge deficits too. Provinces with these characteristics are located
in the Amazonas Region in Brazil, the Caribbean Coast of Nicaragua,
Sipaliwini in Suriname, most of Gabon, and in South Kalimantan and
Sumatera Selatan (Indonesia).

Our results demonstrate that the patterns of under-reporting at a
provincial scale are by no means homogenous. If poorly-researched
provinces were randomly distributed, then adjacent provinces could be
deemed `representative'. However, many under-reported provinces, with
accessibility issues, have faced serious geopolitical challenges in
recent decades. Many of these areas have experienced violent conflicts
including inter-state wars, civil wars, rebel insurgencies, piracy, drug
wars, and other forms of lawlessness. Moreover, poor security impairs
funding and hinders field research (\cite{sutherland_horizon_2012}). For
example, violent conflicts lead to the destruction of habitats, bushmeat
poaching by combatants and displaced communities, poaching for the
illegal wildlife trade, persistent social disruption and a lack of
conservation capacity (\cite{gaynor_war_2016,nackoney_impacts_2014,conteh_assessing_2017,daskin_warfare_2018}). Hence,
knowledge gaps further deepen threats to wildlife populations, which
therefore often go unnoticed.

For example, 7 out of 17 provinces in the Central African Republic (CAR)
are part of our priority list of 100 under-reported areas with
accessibility issues. These include the Mbomou and Haute-Kotto
provinces, where the Chinko nature reserve was established in 2013 (R.
Hickisch, personal communication, January 12, 2018). This reserve covers
19,840 km\textsuperscript{2} and is among the largest managed protected
areas in the entire region. In this area, the accessibility challenge is
considered a consequence of continuing violence and geographic
remoteness, which constrains logistics and increases costs. The
resource-rich provinces of CAR and adjacent countries have experienced
cycles of conflict and civil war for much of the past three decades
(\cite{collier_bottom_2008}). Many rebel groups, including the Lord's Resistance
Army, have sought refuge in remote, ungoverned, and most likely
wildlife-rich provinces (\cite{ondoua_assessment_2017}). Hence, it remains a major
challenge to operate any kind of business in this region, in particular
conservation activities, which require large numbers of staff. In
addition, universities may be reluctant to provide insurance for their
staff to operate in regions for which travel warnings have been issued.

Consequently, a lack of knowledge may constrain long-lasting
conservation planning. For example, the Chinko basin formerly contained
an elephant population numbering thousands of individuals
(Douglas-Hamilton, 1987). Subsequently the population has crashed to a
few hundred, or even fewer, individuals (\cite{hickisch_update_2013}),
due to ivory poaching. The population decline went largely unrecorded.
Since the millennium, major poaching-induced population declines have
occurred in Garamba National Park in the Democratic Republic of Congo
(\cite{canby_elephant_2015}) and in Zakouma National Park in Chad (\cite{fay_ivory_2007}) too.
Again, these declines occurred in under-reported,
accessibility-challenged provinces included on our priority list.
Knowledge deficits and failures to learn from under-reported events are
not limited to elephant populations. The latest IUCN report on African
freshwaters life (\cite{brooks_status_2011}) records a freshwater
species richness of the Chinko drainage system similar to large parts of
the Sahara desert (i.e., presumably 0). Since 2014, however, the Chinko
river has been known among fly-fishing enthusiast as a favoured habitat
for Goliath Tiger fish (F. Botha, personal communication, Apr 10, 2014). \emph{Hydrocynus goliath} is a
very large predatory fish (max. length: 1.5; max. body mass: 50 kg),
documented in the Congo River basin to which the Chinko river is a
tributary via the Oubangui river. A similar pattern applies to the
Pousargues's mongoose \emph{Dologale dybowskii}: Large parts of its
known distribution range overlap with territory of the Lord Resistance
Army. The description of this species is thus based only on a few recent
observations and museum specimen. It has a data deficient status in the
IUCN redlist (\cite{aebischer_dologale_2015}).

However, knowledge gaps are not constrained to the status and trend of
single species. For example, significant changes in the practices of
pastoral communities are occurring in many less accessible provinces in
central Africa (The Economist, 2017, November 9); and these changes have
been largely unrecognised by conservation research. Social changes, the
legacy of violence, and economic factors are leading to larger domestic
stock herd-sizes, and climatic factors combined with increases in
vaccine availability are allowing movement of these herds into areas
formerly unsuitable for grazing (e.g., tse-tse fly habitats). Changing
practices of transhumance -- larger herd sizes, widespread availability
of automatic weapons, and highly potent poisons threaten wildlife in
multiple ways: from bushmeat hunting to the eradication of top predators
(\cite{bouche_has_2010,brugiere_large-scale_2015}). Whilst the
movements of these large herds can be monitored through remote-sensing
methods -- for example as fire cluster data as used herein -- the
subsequent impacts on wildlife (especially terrestrial mammals and
birds) are more difficult to assess remotely, particularly if dense
vegetation appears largely intact. The lack of adequate data on the
response of wildlife to these changes is important in its own right, but
it also represents a missed opportunity to understand and address these
issues before they spread to nearby areas -- as it is occurring with
respect to large-herd shifting pastoralism. Interestingly, although the
priority provinces we identify largely fall outside of `biodiversity
hotspots', as defined by \cite{myers_biodiversity_2000}, they are almost all
adjacent or connected to areas of high conservation concern.

Addressing the systematic under-reporting of biodiversity in less
accessible provinces is by no means a simple issue. Indeed, we do not
propose that scientists should ignore the risks involved in working in
remote provinces -- but we do highlight the consequences of their
current aversion to these risks. However, we do suggest that greater
attention should be afforded to the accessibility bias that we have
demonstrated in existing data sets, and to develop steps to narrow these
gaps. We consider our provincial-scale model a first step for action,
and we have made all the data open accessible ( \cite{raffael_r_2017_rwosconsindex,raffael_r_2017_provincer}) to assist in this
endeavour.

\section*{Acknowledgments}

We thank A. Dickman, E. Dröge, E. Macdonald, H. Bauer, A. Siddig, and A. Ellison for their inputs. M. Strimas-Mackey has helped us with mapping in R, and A. Ruete and D. Williams provided information on further literature and data sets. We thank the audience in the WildCrU colloquium (Oxford, April 2017) and E.J. Milner-Gulland for critical questions and helpful remarks. The results of T. Aebischer's fabulous fieldwork in Central African Republic laid the foundations for this work. This work was begun while R.H. was a student on the Recanati-Kaplan Centre Postgraduate Diploma in International Wildlife Conservation Practice delivered by WildCRU.

\newpage

\section*{Supporting Information}

\subsection*{Data sources}
The data used includes:

\begin{itemize}
\item
  The Global Administrative Areas `GADM' dataset mapping global
  political boundaries~(\cite{gadm_gadm_nodate});
\item
  The NASA~night-time light data (\cite{ngdc_version_2013}) as a proxy for
  spatial~economic activity;
\item
  
  The weekly~fire clusters~of the NASA VIIRs data set~(\cite{nasa_viirs_nodate}) as a
  proxy for human presence (and labelled as a `human presence index')
  
\item
  
  Mean precipitation and temperature data from Worldclim~(\cite{fick_worldclim_2017}) fas a proxy for~possible biodiversity;
  
\item
  
  A~remoteness index constructed using data on: (1) mean travel time to
  major cities~(\cite{nelson_travel_2008}), and (2) Mapbox
  Directions~(\cite{mapbox_introduction_nodate})~travel times from polygonal centre points (centroids) to
  capitals~(\cite{giraud_shortest_2019,rudis_tools_2018}), airports~(\cite{ourairpots_airports_nodate}),
  and seaports~(\cite{national_geospatial_intelligence_agency_world_nodate}). The
  remoteness index is a measure of the~cost and complexity of
  logistics~related to conservation research and action.
  
\item
  
  A dataset on world road area (\cite{ibisch_global_2016}) as a proxy for
  permanent infrastructure.
  
\item
  
  The mean life expectancy~at birth (from 1960 to 2010) as a proxy for
  the level of afflluence (\cite{world_bank_group_life_nodate}).
  
\item
  
  The conservation~knowledge data set that we created from the
  inferred number of Web of Science publications in the Biodiversity and
  Conservation category of SCI and SSCI catalogues (from 1993 to
  2016;~\cite{reuters_web_2012}).
  
\item
  
  The estimated number of~terrestrial mammal species
  present per country, as well as their
  range~(\cite{iucn_iucn_nodate}) and the
  WDPA dataset on protected areas~(\cite{wdpa_protected_nodate}) as control
  indicator for~present biodiversity.
  
\end{itemize}

\hypertarget{data-preparation}{%
\subsection*{Data Preparation}\label{data-preparation}}

The datasets listed above were downloaded from the Internet, and
converted into Feature Collections according to the~OGC GeoPackage
Encoding Standard~(\cite{noauthor_ogc_nodate}). Using the GDAL
command line tool~ogr2ogr~(\cite{noauthor_gdal_2019})~polygonal
datasets were simplified~by removing extraneous bends while preserving
essential shape with a threshold of 0.0002 degrees. As not all polygonal
data had been available in valid form (according to the OGC standard)
the~st\_makevalid function of the Postgis~(\cite{shekhar_postgis_2016})
library~liblwgeom~was used to solve geometry problems.

We initially tried the Natural Earth airport dataset~(\cite{natural_earth_airports_2017}), however, because of this data set being very limited, we opted
instead to use the filtered records for the~large\_airport
and~medium\_airport records from~the ourairports dataset
(\cite{ourairpots_airports_nodate}) instead.~

The world capital city data set was derived using the~hrbrmstr R
Openstreet Map Overpass query library~(\cite{rudis_tools_2018}) query for
administrative centre using the below R code.

\begin{footnotesize}

\begin{verbatim}
library(overpass)
osmcsv <- '
relation["admin_level"="2"]["type"="boundary"]["boundary"="administrative"];
node(r:"admin_centre");
out meta;'
opq <- overpass_query(osmcsv)
head(opq)
write.csv(opq, file = "capital_cities.csv")
\end{verbatim}
\end{footnotesize}

\hypertarget{data-processing}{%
\subsection*{Data Processing}\label{data-processing}}

We extracted spatial data from the above mentioned sets by the level 1
administrative boundary polygons from the GADM v2.8 dataset (also
referred to as `province';~\cite{gadm_gadm_nodate})~ and saved them to file
using country ISO code and province id (all processed data is available
online: ~\cite{hickisch_aggregated_2017,hickisch_data_nodate,hickisch_supplementary_2017}).~
For countries where only level 0 data was available, we used these
instead.

All spatial datasets have been aggregated by averaging (for raster
data), summarising areal extent (for polygons; in sq m) and/or counting
(for points and polygons) and joining non spatial data (e.g. Life
expectancy) using the R script available (\cite{raffael_r_2017_provincer}).

The summary of all province level data, including transformed (log,
division by area) are available (\cite{hickisch_data_nodate});
all prepared input data is available here
\cite{hickisch_supplementary_2017}

Table~\ref{table2} shows the correlation matrix of all model input variables
prepared in the data processing.

\begin{table}[!ht]
\begin{adjustwidth}{-2.25in}{0in} 
\centering

\caption{
{\bf Correlation matrix of model input variables}}
\begin{footnotesize}
\begin{tabular}{@{}rcccccccccccc@{}}

fire\_clusters& 1 & & & & & & & & & & &\\
log\_province\_road\_area\_2 & 0.78** & 1 & & & & & & & & &
&\\
province\_pa\_area & 0.24** & 0.22** & 1 & & & & & & & &
&\\
log\_province\_area\_m2 & 0.82** & 0.92** & 0.27** & 1 & & & & & & &
&\\
province\_pa\_area\_Iab\_II & 0.19** & 0.20** & 0.74** & 0.25** & 1 & &
& & & & &\\
province\_nightlight & -0.35** & -0.38** & -0.10** & -0.51** & -0.09** &
1 & & & & & &\\
Publications & 0.26** & 0.35** & 0.22** & 0.28** & 0.24** & 0.05** & 1 &
& & & &\\
province\_pa\_area\_count & 0.19** & 0.31** & 0.16** & 0.21** & 0.20** &
0.01 & 0.39** & 1 & & & &\\
province\_area\_species\_count & 0.57** & 0.36** & 0.23** & 0.50** &
0.16** & -0.31** & 0.11** & -0.01 & 1 & & &\\
province\_pa\_area\_Iab\_II\_count & 0.13** & 0.18** & 0.28** & 0.15** &
0.57** & -0.05** & 0.28** & 0.32** & 0.06** & 1 & &\\
country\_SP.DYN.LE00.IN & -0.25** & -0.05** & 0.00 & -0.26** & 0.04* &
0.37** & 0.18** & 0.18** & -0.46** & 0.12** & 1 &\\
province\_num\_conflict & 0.02 & 0.05** & -0.01 & 0.06** & -0.01 &
-0.04* & -0.02 & -0.02 & 0.02 & -0.01 & -0.14** & 1\\

\end{tabular}
\end{footnotesize}

\begin{flushleft} 
\end{flushleft}
\label{table2}
\end{adjustwidth}
\end{table}

\hypertarget{indicator-of-road-area}{%
\subsection*{Indicator of Road Area}\label{indicator-of-road-area}}

Road Area is calculated by subtracting summarised roadless area~(Ibisch
et al. 2016) from the GADM administrative polygon~(GADM.org 2017) in
order to represent the known presence of infrastructure (and solve
issues for Greenland and Antarctica, which have no roads).

\textbf{Caveats: not all roads are documented; more roads are documented
in better connected/developed areas}

\hypertarget{indicator-of-human-presence}{%
\subsubsection*{Indicator of Human
Presence}\label{indicator-of-human-presence}}

We decided not to use the Global Human Footprint dataset as a proxy for
human presence. It represents human presence as complex composite
variable of data on roads, villages, cropland, and other land
asset-related variables~(Sanderson et al. 2002). However, global
roadmaps (and road less areas, such as~(Ibisch et al. 2016)), and maps
of villages~(Socioeconomic Data and Applications Center 2017), and even
river data sets~(arcgis.com 2017)~are much better mapped in provinces~of
larger wealth~(Ibisch and Selva 2017). Therefore, a particular drawback
is that it largely underestimates the impact of relatively small groups
of people roaming vast areas, such as transhumant herdsmen~in the Sahel.

\begin{figure}[!h]
\begin{adjustwidth}{-2.25in}{0in} 
\centering
\includegraphics[width=\linewidth]{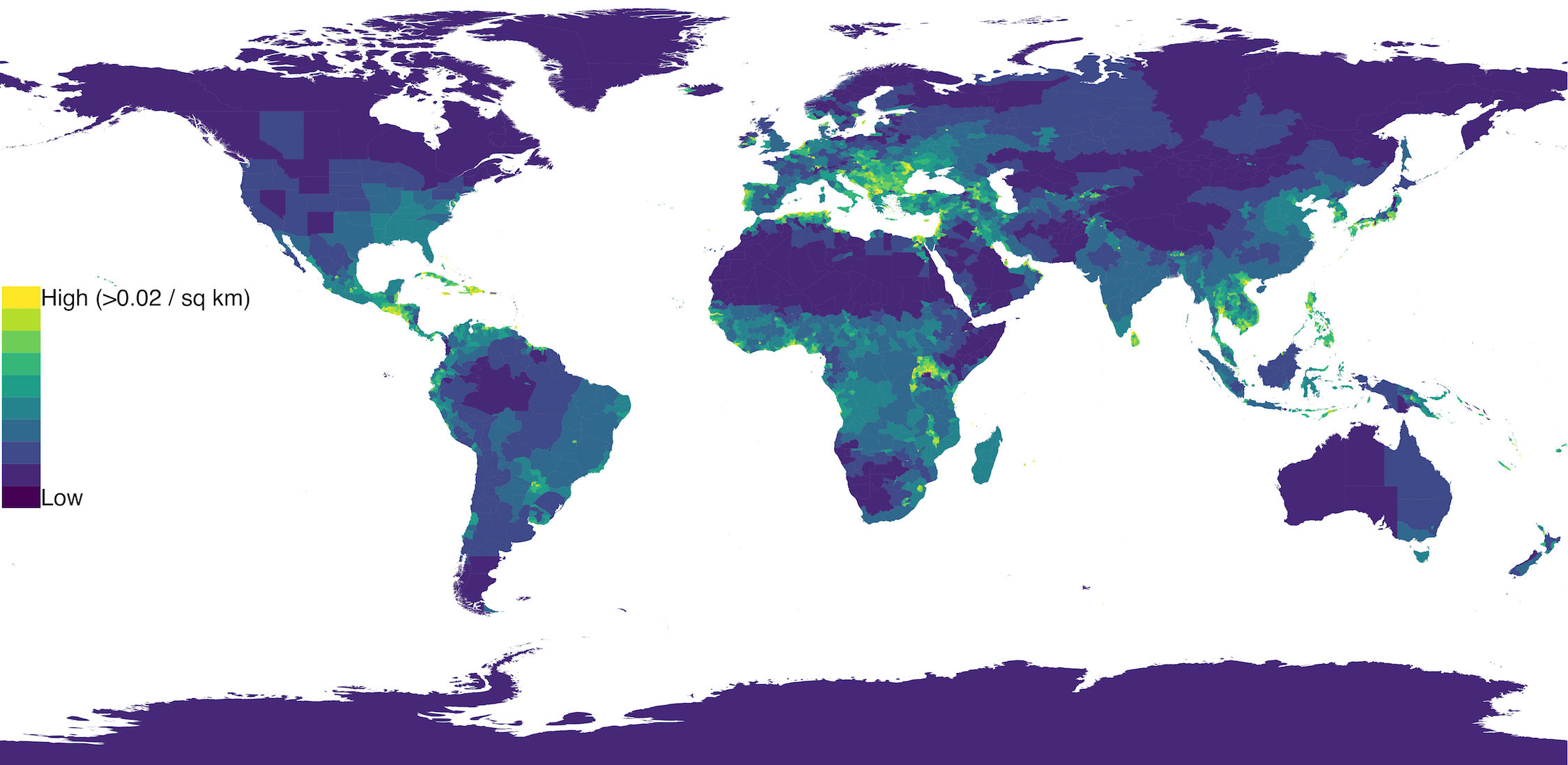}
\caption{{\bf NASA VIIRS fire clusters.}
Map of log transformed frequency of NASA VIIRS fire clusters
per sq km for the time period July 2016 to July 2017. The weekly
clustering of fire points corrects for large wildfires that might
otherwise skew this indicator of human presence. Fire clusters are more
frequent in the underdeveloped regions of the tropics. No correction for
gas flares, and other industrial fires, such as steel production has
been made. This explains the fire cluster frequency in some of northern
areas (data available at \cite{hickisch_aggregated_2017}).}
\label{fig4}
\end{adjustwidth}
\end{figure}

Instead, we opted to use data on fire as a measure of human presence
(see  Fig~\ref{fig4}). The high resolution NASA VIIRS data set on global daily fire points is available to the public from January 2016 onwards. To
correct for large bushfires, eventually resulting in many such
non-independent fire points along a front, we spatially cluster these
points by cutting a weekly dendrogram of geolocations of these fires at
the first level (using the R functions~cutree and~hclust), and subsample
randomly if a province has more than 20,000 fire points in a single
week. The resulting count of fire clusters provides a complementary
indicator of human presence. We tested how well our indices of road
coverage, protected areas and night light predicted the clusters.
Together, these accounted for 61\% of variation among princes in the
frequency of fire clusters. Below the model output.

\begin{footnotesize}

\begin{verbatim}
Call:
lm(formula = log(province_num_weekly_fire_clusters + 1) ~ +log_province_road_area_2 + 
    province_pa_area + province_pa_area_Iab_II + province_nightlight)

Residuals:
    Min      1Q  Median      3Q     Max 
-5.5262 -0.4897  0.1513  0.6531  4.6418 

Coefficients:
                           Estimate Std. Error t value Pr(>|t|)    
(Intercept)              -8.017e+00  1.772e-01 -45.234  < 2e-16 ***
log_province_road_area_2  5.263e-01  8.221e-03  64.014  < 2e-16 ***
province_pa_area          5.443e-12  8.216e-13   6.625 4.00e-11 ***
province_pa_area_Iab_II  -6.999e-12  2.772e-12  -2.524   0.0116 *  
province_nightlight      -7.362e-03  1.281e-03  -5.745 9.93e-09 ***
---
Signif. codes:  0 ‘***’ 0.001 ‘**’ 0.01 ‘*’ 0.05 ‘.’ 0.1 ‘ ’ 1

Residual standard error: 0.9763 on 3595 degrees of freedom
  (1 observation deleted due to missingness)
Multiple R-squared:  0.6117,    Adjusted R-squared:  0.6113 
F-statistic:  1416 on 4 and 3595 DF,  p-value: < 2.2e-16

Analysis of Variance Table

Response: log(province_num_weekly_fire_clusters + 1)
                           Df Sum Sq Mean Sq   F value    Pr(>F)    
log_province_road_area_2    1 5312.3  5312.3 5573.9008 < 2.2e-16 ***
province_pa_area            1   47.9    47.9   50.2485 1.623e-12 ***
province_pa_area_Iab_II     1    5.9     5.9    6.1841   0.01294 *  
province_nightlight         1   31.5    31.5   33.0094 9.934e-09 ***
Residuals                3595 3426.3     1.0                        
---
Signif. codes:  0 ‘***’ 0.001 ‘**’ 0.01 ‘*’ 0.05 ‘.’ 0.1 ‘ ’ 1

\end{verbatim}
\end{footnotesize}

\textbf{Caveats:~ industrial fires are mixed in (no correction for gas
flares). We presume the artisanal fires to dominate}

\hypertarget{indicator-of-remoteness}{%
\subsubsection*{Indicator of Remoteness}\label{indicator-of-remoteness}}

As an indicator of remoteness, we propose to extend the travel time to
major cities dataset~(Nelson 2008) by the Mapbox Directions~(mapbox.com
2017) mean travel time from the province centre point (centroid) to
capitals~(\cite{rudis_tools_2018}), airports~(\cite{natural_earth_airports_2017,ourairpots_airports_nodate}), and
seaports~(\cite{national_geospatial_intelligence_agency_world_nodate}), as more remote
places are likely understudied.

\begin{figure}[!h]
\begin{adjustwidth}{-2.25in}{0in}
\includegraphics[width=\linewidth]{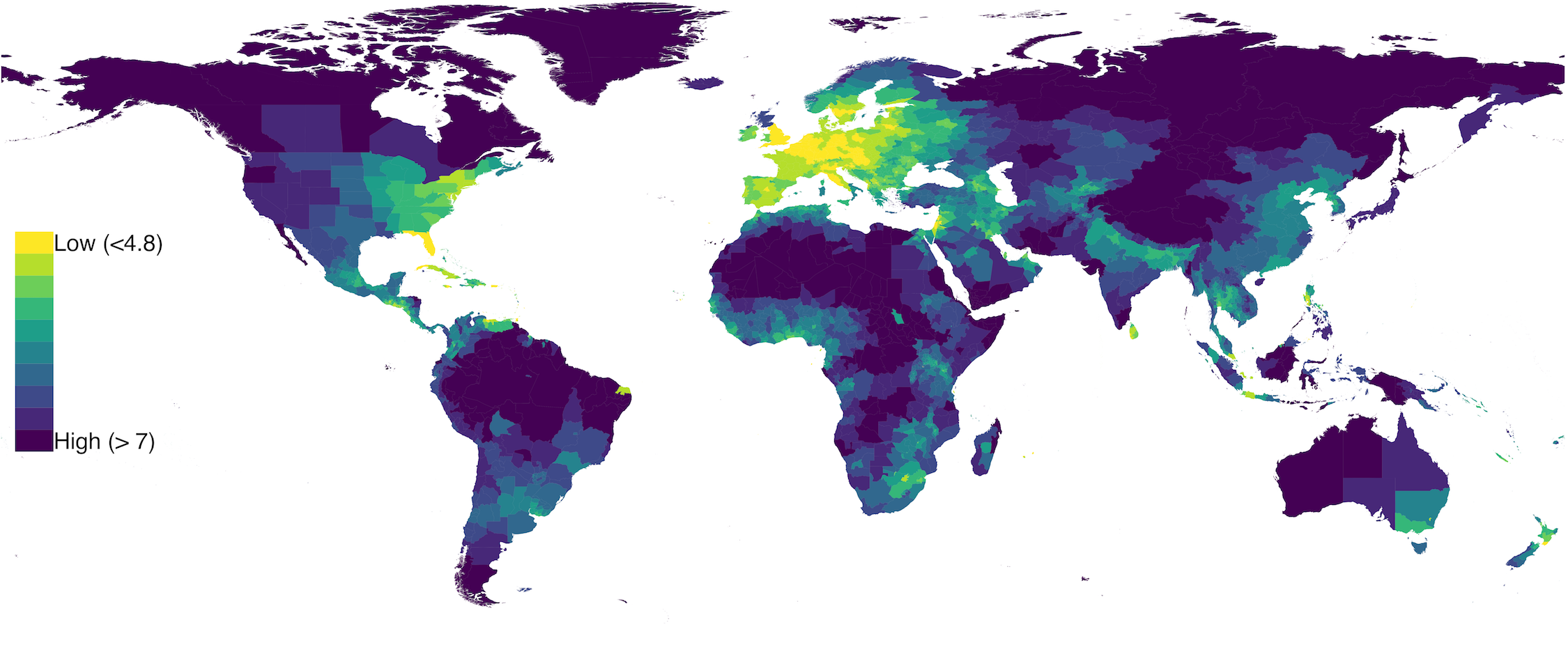}
\caption{{\bf  Log transformed average travel times.}
Map illustrating log transformed average travel time from the
province central point to the closest settlement over 50,000
inhabitants, to the three spatially closest capital cities, ports and
airports. For large neighboring states, such as Russia, China, India,
Brazil, Canada, and the United States of America the distance to the
capital city has a very strong impact, representing the governance
aspects of remoteness (data available at \cite{hickisch_aggregated_2017}).}
\label{fig5}
\end{adjustwidth}
\end{figure}

The Remoteness Index is the log transformed average (omitting~NAs) of
the~mean travel time from the province centre point (the centroid as
calculated by R package sf function st\_centroid) to the closest
settlement over 50,000 inhabitants, and the travel times to the three
spatially closest capital cities (legislative centres), airports
(accessibility), and seaports (supply chain)~by linking
the~osrm~R-package~(\cite{giraud_shortest_2019})~to the Mapbox Distance API~(\cite{mapbox_introduction_nodate}). In order to validate this remoteness index, we test the
hypothesis that larger areas with less nightlight, and less roads are
more remote (R\textsuperscript{2}= 0.45, 3595 df; see model output
below; see Fig~\ref{fig5}).

\begin{footnotesize}
\begin{verbatim}
Call:
lm(formula = log(province_remoteness_min + 1) ~ +log_province_area_m2 + 
    log_province_road_area_2 + province_nightlight)

Residuals:
    Min      1Q  Median      3Q     Max 
-2.5618 -0.4290 -0.0708  0.3650  3.3375 

Coefficients:
                           Estimate Std. Error t value Pr(>|t|)    
(Intercept)               4.9857188  0.1199225   41.57   <2e-16 ***
log_province_area_m2      0.3785360  0.0118679   31.90   <2e-16 ***
log_province_road_area_2 -0.3409079  0.0135638  -25.13   <2e-16 ***
province_nightlight      -0.0184455  0.0009732  -18.95   <2e-16 ***
---
Signif. codes:  0 ‘***’ 0.001 ‘**’ 0.01 ‘*’ 0.05 ‘.’ 0.1 ‘ ’ 1

Residual standard error: 0.6698 on 3595 degrees of freedom
  (2 observations deleted due to missingness)
Multiple R-squared:  0.4529,    Adjusted R-squared:  0.4524 
F-statistic:   992 on 3 and 3595 DF,  p-value: < 2.2e-16

Analysis of Variance Table

Response: log(province_remoteness_min + 1)
                           Df  Sum Sq Mean Sq F value    Pr(>F)    
log_province_area_m2        1  746.93  746.93 1664.77 < 2.2e-16 ***
log_province_road_area_2    1  427.08  427.08  951.89 < 2.2e-16 ***
province_nightlight         1  161.18  161.18  359.25 < 2.2e-16 ***
Residuals                3595 1612.95    0.45                      
---
Signif. codes:  0 ‘***’ 0.001 ‘**’ 0.01 ‘*’ 0.05 ‘.’ 0.1 ‘ ’ 1
\end{verbatim}
\end{footnotesize}

\textbf{Caveats: the travel times for low traffic reported roads/little
driven roads, (etc in Central Africa) are very optimistic (eg 14 hours
instead of 40 from Chinko to Bangui) because the political condition
(road blocks) are not documented.}

\hypertarget{indicator-of-scientific-knowledgepublications}{%
\subsubsection*{Indicator of Scientific
Knowledge/Publications}\label{indicator-of-scientific-knowledgepublications}}

As an indicator for scientific field sampling effort, we query the Web
of Science SCI and SSCI catalogues for country and province names in
various languages. We used the R-package~rwos~(\cite{barnier_r_2019}) to query
the Web of Science Lite API~(\cite{reuters_web_2012});~country~and~province~name
were taken from the~GADM v.2.8~dataset~(\cite{gadm_gadm_nodate}) and the
R-package~countrycode~(\cite{arel-bundock_r_2019}),~and results were filtered for
the Web of Science category biodiversity and conservation.

This method is not aiming to quantify the actual number of publications
per province, but works as a relative index of scientific attention. The
scope of these queries is 1993 to 2016 - starting with the United
Nations Earth Summit conference in Rio de Janeiro.

All queries and the result total have been retrieved and stored with the
help of the R script available
in ~\cite{raffael_r_2017_rwosconsindex}~and
allows for reproduction for any institution that holds Web of Science
access.

To obtain information on those provinces, that have only very few
mentions since 1993, we infer a conservation publication estimate, based
on the proportion of~biodiversity and conservation results to those for
the country overall, and multiplied by the overall number of
publications for that province. If a country has a total of N
publications and C conservation publications, and x publications for a
province, the inferred publication density (IPD) is calculated as
IPD=(C/N)*x.

We note that countries appear to have a reputation for a certain
research topic that appears to be rather stable over time (a certain
inertia effect). We do use the IPD instead of simply the number of
biodiversity and conservation category WoS publications by province, as
we'd then have to neglect the detail within the more underrepresented
provinces.

We checked the inferred result by multiplying the provincial result
across all categories, which correlates strongly with the actual number
of conservation related publications (R\textsuperscript{2}=0.98, 3596
df; see below). ~

\begin{footnotesize}
\begin{verbatim}
Call:
lm(formula = province_wos_cons_inferred ~ province_wos_cons)

Residuals:
    Min      1Q  Median      3Q     Max 
-797.80    0.95    1.00    1.29  368.07 

Coefficients:
                   Estimate Std. Error t value Pr(>|t|)    
(Intercept)       -0.948379   0.352278  -2.692  0.00713 ** 
province_wos_cons  0.891525   0.002339 381.235  < 2e-16 ***
---
Signif. codes:  0 ‘***’ 0.001 ‘**’ 0.01 ‘*’ 0.05 ‘.’ 0.1 ‘ ’ 1

Residual standard error: 21.03 on 3596 degrees of freedom
  (3 observations deleted due to missingness)
Multiple R-squared:  0.9759,    Adjusted R-squared:  0.9758 
F-statistic: 1.453e+05 on 1 and 3596 DF,  p-value: < 2.2e-16

Analysis of Variance Table

Response: province_wos_cons_inferred
                    Df   Sum Sq  Mean Sq F value    Pr(>F)    
province_wos_cons    1 64248196 64248196  145340 < 2.2e-16 ***
Residuals         3596  1589630      442                      
---
Signif. codes:  0 ‘***’ 0.001 ‘**’ 0.01 ‘*’ 0.05 ‘.’ 0.1 ‘ ’ 1

\end{verbatim}
\end{footnotesize}

In a linear mixed model correcting for country as random effect (see
details below), this inferred conservation index is positively explained
most strongly by the countries' life expectancy at birth (estim.
=1.446e-02 , p = 0.015), the nightlight index (estim. =1.699e-02 , p =
 2e-16), and the logistic area of buffered roads (estim.
=1.415e-01 , p =  2e-16).

\begin{footnotesize}
\begin{verbatim}
Linear mixed model fit by REML t-tests use Satterthwaite approximations to
  degrees of freedom [lmerMod]
Formula: log(province_wos_cons_inferred + 1) ~ +province_pa_area_count +  
    province_area_species_count + province_pa_area_Iab_II_count +  
    province_nightlight + country_SP.DYN.LE00.IN + log_province_road_area_2 +  
    (1 | country_iso)

REML criterion at convergence: 6932.5

Scaled residuals: 
    Min      1Q  Median      3Q     Max 
-5.2709 -0.4094 -0.1212  0.1968  7.4844 

Random effects:
 Groups      Name        Variance Std.Dev.
 country_iso (Intercept) 0.6202   0.7875  
 Residual                0.3611   0.6009  
Number of obs: 3428, groups:  country_iso, 204

Fixed effects:
                                Estimate Std. Error         df t value Pr(>|t|)
(Intercept)                   -3.496e+00  4.569e-01  3.460e+02  -7.652 1.98e-13
province_pa_area_count         6.530e-04  1.132e-04  3.314e+03   5.770 8.65e-09
province_area_species_count    1.521e-03  4.132e-04  3.290e+03   3.680 0.000237
province_pa_area_Iab_II_count  3.339e-03  5.226e-04  3.333e+03   6.388 1.91e-10
province_nightlight            1.699e-02  1.195e-03  3.411e+03  14.213  < 2e-16
country_SP.DYN.LE00.IN         1.446e-02  5.912e-03  2.020e+02   2.445 0.015330
log_province_road_area_2       1.415e-01  1.188e-02  3.271e+03  11.910  < 2e-16
                                 
(Intercept)                   ***
province_pa_area_count        ***
province_area_species_count   ***
province_pa_area_Iab_II_count ***
province_nightlight           ***
country_SP.DYN.LE00.IN        *  
log_province_road_area_2      ***
---
Signif. codes:  0 ‘***’ 0.001 ‘**’ 0.01 ‘*’ 0.05 ‘.’ 0.1 ‘ ’ 1

Correlation of Fixed Effects:
            (Intr) prvnc_p__ prvnc_r__ p___I_ prvnc_ c_SP.D
prvnc_p_r_c  0.101                                         
prvnc_r_sp_ -0.086  0.014                                  
prv___I_II_  0.097 -0.195    -0.118                        
prvnc_nghtl -0.176 -0.037     0.076    -0.036              
c_SP.DYN.LE -0.831 -0.036     0.185    -0.044 -0.091       
lg_prvn___2 -0.502 -0.142    -0.259    -0.097  0.406 -0.042
Analysis of Variance Table of type III  with  Satterthwaite 
approximation for degrees of freedom
                              Sum Sq Mean Sq NumDF  DenDF F.value    Pr(>F)    
province_pa_area_count        12.023  12.023     1 3314.3  33.294 8.651e-09 ***
province_area_species_count    4.892   4.892     1 3290.0  13.546 0.0002365 ***
province_pa_area_Iab_II_count 14.738  14.738     1 3333.1  40.811 1.909e-10 ***
province_nightlight           72.949  72.949     1 3411.4 202.004 < 2.2e-16 ***
country_SP.DYN.LE00.IN         2.159   2.159     1  202.1   5.979 0.0153298 *  
log_province_road_area_2      51.229  51.229     1 3270.7 141.860 < 2.2e-16 ***
---
Signif. codes:  0 ‘***’ 0.001 ‘**’ 0.01 ‘*’ 0.05 ‘.’ 0.1 ‘ ’ 1
\end{verbatim}
\end{footnotesize}

The number of protected areas in the provinces is a significant
predictor of publication density. It is plausible that this is because
this represents research that is done in reserves, and also covers a
possible bias that more research being done in protected areas, as
highlighted in~(Martin, Blossey, and Ellis 2012; Meyer, Weigelt, and
Kreft 2016). The fact that the number of species present according to
IUCN species range maps (estim.= 8.135e-02, p = 0.0002) also explains
the number of publications might, however, reflect a causal relationship
in the other direction -- i.e that range maps are themselves an artefact
of a spatial publication frequency bias.

For model explanation, and statistical analysis, the R-packages lme4,
lmerTest and vegan (Bates et al. 2015; Kuznetsova, Bruun Brockhoff, and
Haubo Bojesen Christensen 2016; Oksanen et al. 2017) were used.

\textbf{Caveats: SCI + SSCI is dominated by Western literatures; also we
use the former country names, and the current name in seven languages
(``de'',``ar'',``fr'',``en'',``es'',``ru'',``zh''), but we do not
correct for changes in the province names since 1993.}

\hypertarget{principal-component-analysis}{%
\subsubsection*{Principal Component
Analysis}\label{principal-component-analysis}}

Having this dataset (n=3638) on the province level available, we use the
R built-in function~prcomp to run a principal component analysis, and
characterise the provinces with the variables defined earlier (see Table~\ref{table3}).

\begin{table}[!ht]
\begin{adjustwidth}{-2.25in}{0in} 
\centering
\caption{
{\bf Attributes used in the Principal Component Analysis}}

\begin{footnotesize}

\begin{tabular}[]{@{}lllll@{}}

\textbf{Attribute} & \textbf{Timespan} & \textbf{Aggregation Method} & \textbf{Reason} &
\textbf{Sources}\\ \thickhline

Province.Area & 2017 & Sum & To correct for polygon size & \cite{gadm_gadm_nodate}; levels: 0, 1\\
Temperature & 1970-2000 & Mean (omitting Nas) & Possible biodiversity &
\cite{fick_worldclim_2017}; band: 1\\
Precipitation & 1970-2000 & Mean (omitting Nas) & Possible biodiversity
& \cite{fick_worldclim_2017}; band: 12\\
Publications & 1993-2016 & Log-transformed sum & Scientific Knowledge & \cite{reuters_web_2012}; catalogues: SCI, SSCI\\
Fire.Clusters & Jul 2016-Jun 17 (1 year) & Log-transformed sum & Human
Presence &
\cite{nasa_viirs_nodate}\\
Remoteness & 2008,2017 &
Log-tr. mean (omit. Nas) & Logistic Cost \& Complexity & \cite{nelson_travel_2008,mapbox_introduction_nodate}\\
Road.Area & 2016 & Log-transformed sum & Infrastructure & \cite{gadm_gadm_nodate,ibisch_global_2016}\\
Life.Expectancy & 1960-2010 & Mean (omitting Nas) & Development & \cite{world_bank_group_life_nodate}\\
Night.Light & 2013 & Mean (omitting Nas) & Economic Activity & \cite{ngdc_version_2013}\\

\end{tabular}
\end{footnotesize}

\begin{flushleft} 
\end{flushleft}
\label{table3}
\end{adjustwidth}
\end{table}

All records are complete on spatially derived attributes, however, the
other data columns were often found to be incomplete, which resulted in~
reduction of characterised provinces (see run\_pca~R script available at \cite{raffael_r_2017_provincer};
it uses the~ggfortify~R-package~(\cite{tang_ggfortify:_2016}
graph).~

\hypertarget{pca-results}{%
\subsubsection*{PCA Results}\label{pca-results}}

Running the PCA on the attributes illustrated below the synthetic
variables PC1 and PC2 explain more than~54\% of the variance (see
below). All PCA output and graphics are available at
\cite{hickisch_supplementary_2017}

\begin{footnotesize}

\begin{verbatim}
Importance of components:
                          PC1    PC2    PC3    PC4     PC5     PC6     PC7
Standard deviation     1.7192 1.3840 1.0873 0.8879 0.86038 0.73650 0.60917
Proportion of Variance 0.3284 0.2128 0.1314 0.0876 0.08225 0.06027 0.04123
Cumulative Proportion  0.3284 0.5412 0.6726 0.7602 0.84242 0.90270 0.94393
                           PC8     PC9
Standard deviation     0.57678 0.41471
Proportion of Variance 0.03696 0.01911
Cumulative Proportion  0.98089 1.00000
\end{verbatim}
\end{footnotesize}

Provinces, that had to be dropped due to incomplete information come
from 53 countries (n=184; see Table~\ref{table5}) and were mainly incomplete for
life expectancy (170 NAs), as well as temperature (27 NAs),
precipitation (24 NAs), remoteness (3 NAs), and inferred publications (3
NAs). Most of these provinces are ex-colonial small islands where the
World Bank datasets do not record Life Expectancy. There is also no life
expectancy data for Antarctica, Western Sahara, and Kosovo.

\begin{table}[!ht]
\begin{adjustwidth}{-2.25in}{0in} 
\centering
\caption{
{\bf Provinces that were dropped before the PCA as of NA values are
largely formerly colonial islands, where Life expectancy is documented
together with the mainland counterpart. Table not included here due to volume;  available at
\cite{hickisch_supplementary_2017} in 
\href{https://zenodo.org/api/files/c8944018-9d5a-45ff-a695-705eb8ec4b9d/dropped_provinces.csv?versionId=adb2de9e-d825-4e0c-a2f7-dca6e471852f}{{dropped\_provinces.csv}}}}

\label{table5}
\end{adjustwidth}
\end{table}

\begin{table}[!ht]
\begin{adjustwidth}{-2.25in}{0in} 
\centering
\caption{
{\bf Scaled input to the PCA (without NAs) -- first 20 rows. All
3602 records available at \cite{hickisch_supplementary_2017}
in
\href{https://zenodo.org/api/files/c8944018-9d5a-45ff-a695-705eb8ec4b9d/pca_input_scaled.csv?versionId=1fdf843a-f72e-48a1-8539-1acd6488ad92}{{pca\_input\_scaled.csv}}}}

\begin{tabular}{@{}rrrrrrrrrl@{}}

\textbf{Area} & \textbf{Temp} & \textbf{Precip} & \textbf{Public.} & \textbf{Road} &
\textbf{Fire} & \textbf{Life} & \textbf{Light} & \textbf{Remote} &
\textbf{Name}\\ \thickhline

-0.14 & 1.03 & 1.96 & 0.46 & 0.36 & 0.1 & -0.79 & -0.58 & 3.63 & Andaman
and Nicobar IND 1\\
0.49 & 1.13 & -0.42 & 1.92 & 1.47 & 1.8 & -0.79 & -0.16 & 0.64 & Andhra
Pradesh IND 2\\
0.17 & -0.44 & 0.8 & 0.77 & 0.7 & 1.12 & -0.79 & -0.63 & 1.16 &
Arunachal Pradesh IND 3\\
0.15 & 0.69 & 1.48 & 1.4 & 1.22 & 1.37 & -0.79 & -0.43 & 0.33 & Assam
IND 4\\
0.22 & 0.9 & -0.08 & 1.21 & 1.42 & 1.21 & -0.79 & -0.43 & 0.09 & Bihar
IND 5\\
-0.17 & 0.72 & -0.26 & 0.58 & -1.12 & -1.02 & -0.79 & 3.85 & 0.42 &
Chandigarh IND 6\\
0.39 & 0.91 & 0.18 & 0.41 & 1.11 & 1.62 & -0.79 & -0.36 & 0.56 &
Chhattisgarh IND 7\\
-0.16 & 0.99 & 1.43 & -0.62 & -1.14 & -0.27 & -0.79 & 0.91 & 0.44 &
Dadra and Nagar Haveli IND 8\\
-0.17 & 1.02 & 0.4 & -0.63 & -1.27 & -0.76 & -0.79 & 2.75 & 1.06 & Daman
and Diu IND 9\\
-0.15 & 1.06 & 2.08 & 0.91 & 0.23 & 0.48 & -0.79 & 0.32 & 0.4 & Goa IND
10\\
0.59 & 1.05 & -0.63 & 1.74 & 1.7 & 1.75 & -0.79 & -0.19 & 0.95 & Gujarat
IND 11\\
0.01 & 0.82 & -0.77 & 0.99 & 1.3 & 1.17 & -0.79 & 0.57 & 0.38 & Haryana
IND 12\\
0.06 & -1.16 & -0.12 & 1.18 & 0.97 & 0.9 & -0.79 & -0.44 & 1.07 &
Himachal Pradesh IND 13\\
0.26 & -1.96 & -0.92 & 0.49 & 1.11 & 0.81 & -0.79 & -0.55 & 1.46 & Jammu
and Kashmir IND 14\\
0.16 & 0.84 & 0.05 & 0.59 & 1.04 & 1.28 & -0.79 & -0.44 & 0.03 &
Jharkhand IND 15\\
0.61 & 0.87 & -0.04 & 1.92 & 1.99 & 1.78 & -0.79 & -0.15 & 0.65 &
Karnataka IND 16\\
-0.01 & 0.94 & 1.98 & 2.03 & 1.24 & 0.92 & -0.79 & 0.02 & 0.79 & Kerala
IND 17\\
-0.17 & 1.15 & 0.77 & 0.02 & -1.67 & -1.91 & -0.79 & -0.29 & 0.81 &
Lakshadweep IND 18\\
1.09 & 0.87 & -0.18 & 1.16 & 1.63 & 2.08 & -0.79 & -0.44 & 0.49 & Madhya
Pradesh IND 19\\

\end{tabular}

\begin{flushleft} 
\end{flushleft}
\label{table6}
\end{adjustwidth}
\end{table}

The rotational matrix resulting from~the PCA is indicated in the block
below.
\begin{footnotesize}

\begin{verbatim}
                             PC1         PC2         PC3         PC4
Province.Area                0.3318507  0.16119946 -0.23316526  0.60370130
Temperature                 -0.1528287 -0.49549426 -0.43369167 -0.25613610
Precipitation               -0.2054726 -0.28692870 -0.58012590  0.29654050
Publications                 0.2387599  0.31759411 -0.55359429 -0.06113565
Road.Area                    0.4963806  0.17453185 -0.04811661 -0.28960205
Fire.Clusters                0.4672544 -0.03833896 -0.20032359 -0.43068460
Life.Expectancy             -0.1792966  0.56012284 -0.11065538  0.16897863
Night.Light                 -0.3526410  0.31002747 -0.21099537 -0.18497600
log_province_remoteness_min  0.3821910 -0.32174555  0.12379177  0.38227835
                             PC5        PC6        PC7         PC8
Province.Area               -0.471584229  0.3927448 -0.2447083 -0.08491258
Temperature                 -0.269149264  0.1416724 -0.1689974  0.55908783
Precipitation                0.482472067  0.1330648  0.2387677 -0.37468527
Publications                 0.012939285 -0.6776335 -0.2358452  0.04205512
Road.Area                    0.204684559  0.1506349  0.1018911 -0.06788357
Fire.Clusters                0.003254859  0.3674552  0.2494079 -0.05224260
Life.Expectancy              0.341788717  0.2659446  0.1906336  0.62006840
Night.Light                 -0.555636938 -0.0381395  0.5729529 -0.20694841
log_province_remoteness_min -0.069007042 -0.3448068  0.6014337  0.32156342
                             PC9
Province.Area               -0.03655290
Temperature                 -0.20924377
Precipitation               -0.05555159
Publications                 0.12319613
Road.Area                   -0.74658140
Fire.Clusters                0.59549747
Life.Expectancy              0.07031861
Night.Light                 -0.13965107
log_province_remoteness_min -0.01301993
\end{verbatim}
\end{footnotesize}

An excerpt of principal component coordinates for the lower right
quadrant of the PC1 and PC2 plot, are indicated in Table~\ref{table7}. The score
used to rank the provinces is composed of the absolute value of the PC1
plus the absolute value of PC2 coordinates.

\begin{table}[!ht]
\centering
\caption{
{\bf Principal component coordinates for the lower right quadrant of
the PC1 and PC2 plot. Score is the sum of absolute PC1 plus absolute PC2
-- first 20 rows. All records are available at
\cite{hickisch_supplementary_2017} in
\href{https://zenodo.org/api/files/c8944018-9d5a-45ff-a695-705eb8ec4b9d/pca_bottom_right.csv?versionId=45ecf201-933a-4e9a-9287-ca35b89a3196}{{pca\_bottom\_right.csv}}}}

\begin{footnotesize}

\begin{tabular}[]{@{}lccccccccc@{}}

& \textbf{PC1} & \textbf{PC2} & \textbf{PC3} & \textbf{PC4} & \textbf{PC5} & \textbf{PC6} & \textbf{PC7} & \textbf{PC8} & \textbf{PC9}\\ \thickhline

Andaman and Nicobar IND 1 & 1.35 & -2.71 & -1.18 & 1.42 & 0.64 & -1.4 &
1.98 & 0.66 & -0.5\\
Andhra Pradesh IND 2 & 2.71 & -0.18 & -1.8 & -1.09 & -0.81 & -0.42 &
-0.22 & 0.34 & -0.09\\
Arunachal Pradesh IND 3 & 1.83 & -0.64 & -0.69 & 0.25 & 0.5 & -0.46 &
0.54 & -0.64 & 0.3\\
Assam IND 4 & 1.69 & -0.78 & -2.16 & -0.53 & 0.61 & -0.19 & 0.1 & -0.59
& -0.17\\
Bihar IND 5 & 1.9 & -0.36 & -1.27 & -1.05 & -0.19 & -0.14 & -0.46 & 0.01
& -0.43\\
Chhattisgarh IND 7 & 1.94 & -0.86 & -1.07 & -0.66 & -0.36 & 0.51 & 0.11
& -0.0 & -0.05\\
Gujarat IND 11 & 3.0 & -0.17 & -1.54 & -0.93 & -0.93 & -0.35 & -0.08 &
0.45 & -0.3\\
Haryana IND 12 & 1.55 & -0.05 & -0.78 & -1.46 & -0.94 & -0.41 & 0.25 &
0.15 & -0.45\\
Jharkhand IND 15 & 1.51 & -0.65 & -0.94 & -0.95 & -0.15 & 0.24 & -0.32 &
-0.07 & -0.13\\
Karnataka IND 16 & 3.01 & -0.02 & -1.99 & -0.92 & -0.56 & -0.24 & -0.08
& 0.01 & -0.5\\
Kerala IND 17 & 1.4 & -0.88 & -2.8 & -0.37 & 0.66 & -0.99 & 0.51 & -0.51
& -0.53\\
Madhya Pradesh IND 19 & 3.13 & -0.21 & -1.67 & -0.44 & -0.94 & 0.73 &
-0.35 & -0.03 & -0.1\\
Maharashtra IND 20 & 3.22 & -0.09 & -2.11 & -0.54 & -0.97 & 0.5 & -0.21
& -0.08 & -0.3\\
Manipur IND 21 & 0.85 & -0.91 & -0.76 & -0.03 & 0.54 & -0.5 & 0.04 &
-0.41 & -0.16\\
Meghalaya IND 22 & 0.44 & -1.57 & -2.47 & 0.39 & 1.68 & -0.06 & 0.53 &
-1.33 & -0.09\\
Mizoram IND 23 & 0.75 & -1.45 & -1.09 & 0.15 & 0.82 & -0.17 & 0.38 &
-0.44 & -0.33\\
Nagaland IND 24 & 0.67 & -1.19 & -0.45 & 0.17 & 0.53 & -0.29 & 0.29 &
-0.38 & -0.02\\
Odisha IND 26 & 2.19 & -0.74 & -1.33 & -0.74 & -0.3 & 0.55 & -0.03 &
-0.06 & -0.15\\
Tamil Nadu IND 31 & 2.81 & -0.16 & -1.88 & -1.19 & -0.75 & -0.66 & 0.21
& 0.36 & -0.47\\

\end{tabular}
\end{footnotesize}

\begin{flushleft} 
\end{flushleft}
\label{table7}
\end{table}

The top 100 ranking provinces are aggregated by~country in Table~\ref{table8} and
in detail listed in Table~\ref{table9}. We manually checked all provinces omitted
from the PCA to confirm that they would not have made the priority list
in terms of remoteness, area, fire and publications.

\begin{table}[!ht]
\centering
\caption{
{\bf Top 100 provinces according to PCA lower left quadrant score
aggregated by country.}}
\begin{tabular}[]{@{}lc@{}}

\textbf{Country} & \textbf{Number of Provinces in Top 100}\\ \thickhline

Angola & 12\\
South Sudan & 10\\
Chad & 8\\
Ethiopia & 8\\
Mozambique & 8\\
Somalia & 8\\
Central African Republic & 7\\
Democratic Republic of the Congo & 7\\
Mali & 7\\
Niger & 5\\
Zambia & 5\\
Brazil & 3\\
Sudan & 3\\
Bolivia & 2\\
Madagascar & 2\\
Eritrea & 1\\
India & 1\\
Nigeria & 1\\
Peru & 1\\
Tanzania & 1\\

\end{tabular}

\begin{flushleft} 
\end{flushleft}
\label{table8}
\end{table}

\begin{table}[!ht]
\begin{adjustwidth}{-2.25in}{0in} 
\centering
\caption{
{\bf Top 100 provinces according to PCA lower right quadrant score - first 20 rows. All records are available at \cite{hickisch_supplementary_2017} in
\href{https://zenodo.org/api/files/c8944018-9d5a-45ff-a695-705eb8ec4b9d/top100\%20detail.csv?versionId=c2d28cb1-a96a-4af6-8e0c-34a22597428f}{{top100
detail.csv}}}}

\begin{tabular}[]{@{}lrlccccc@{}}

\textbf{Country}  & \textbf{ID} & \textbf{Province}  & \textbf{Area (m\textsuperscript{2})} & \textbf{Temp.} & \textbf{Precip.} &
\textbf{Pub.} & \textbf{Road.Area}\\ \thickhline

Tanzania & 22 & Ruvuma & 63075882644.83 & 296.13 & 1184.87 & 0.45 &
22.81\\
Central African Republic & 1 & Bamingui-Bangoran & 58468947111.12 &
299.4 & 1152.17 & 0 & 21.15\\
Central African Republic & 4 & Haut-Mbomou & 56355703351.32 & 298.12 &
1435.35 & 0 & 21.1\\
Central African Republic & 5 & Haute-Kotto & 86049245472.47 & 297.62 &
1316.04 & 0.01 & 21.6\\
Central African Republic & 9 & Mbomou & 60150494915.11 & 297.75 & 1618.5
& 0.01 & 21.55\\
Madagascar & 4 & Mahajanga & 151672635669.44 & 297.98 & 1486.65 & 1.75 &
22.87\\
Central African Republic & 13 & Ouaka & 49209480500.37 & 297.66 &
1451.32 & 0.01 & 21.78\\
Madagascar & 6 & Toliary & 164030931907.8 & 297.16 & 832.04 & 0.82 &
23.9\\
Central African Republic & 15 & Ouham & 52964954181.79 & 298.75 &
1330.14 & 0.04 & 22.42\\
Sudan & 8 & North Darfur & 317200633021.61 & 298.54 & 147.53 & 0.12 &
21.81\\
Central African Republic & 17 & Vakaga & 46696061043.21 & 299.37 &
886.35 & 0 & 21.91\\
Chad & 2 & Batha & 90412921603.27 & 302.74 & 290.92 & 0 &
21.56\\
Chad & 3 & Borkou & 256248321053.96 & 299.67 & 30.26 & 0.04 &
21.16\\
Chad & 4 & Chari-Baguirmi & 46030941104.07 & 301.6 & 690.07 & 0.04 &
21.3\\
Sudan & 14 & South Darfur & 78133459700.04 & 299.37 & 651.31 & 0.22 &
21.87\\
Chad & 7 & Guéra & 61038076376.23 & 301.65 & 734.91 & 0.05 &
21.96\\
Peru & 17 & Loreto & 375548002886.75 & 299.26 & 2525.32 & 1.44 &
21.65\\
Sudan & 17 & West Kurdufan & 112959212480.54 & 300.43 & 491.84 & 0.11 &
22.66\\
Chad & 10 & Lac & 19842923740.3 & 300.96 & 262.83 & 0.02 &
22.3\\

\end{tabular}

\begin{flushleft} 
\end{flushleft}
\label{table9}
\end{adjustwidth}
\end{table}


\printbibliography

\end{document}